\documentclass{ws-procs9x6}

\begin{document}

\title{Observability and Geometry \\ in \\ Three dimensional quantum gravity}

\author{Karim Noui and Alejandro Perez}

\address{Center for Gravitational Physics and Geometry, 
Pennsylvania State University \\
University Park, PA 16802, USA }


\maketitle

\abstracts{We consider the coupling between massive and spinning particles and three dimensional gravity. This allows us to construct geometric operators (distances between particles) as Dirac observables. We quantize the system \`a la loop quantum gravity: we give a description of the kinematical Hilbert space and construct the associated spin-foam model. We quantize the physical distance operator and compute its spectrum.}

\section{Coupling gravity to point particles}
It is well known \cite{Witten} that three dimensional gravity can be formulated as a Chern-Simons theory whose gauge group $G$ depends on the signature of the metric and on the sign of the cosmological constant. In this article, we will exclusively concentrate on the euclidean case with zero cosmological constant. In that case, the gauge group is the isometry group $G=ISU(2)$ of the three dimensional euclidean space $\mathbb E^3$. We will denote by $(P_a)_{a=0,1,2}$ and $(J_a)_{a=0,1,2}$ respectively the translational and rotational generators of the Lie algebra $\mathfrak g=isu(2)$.

The coupling between a point particle (mass $m$ and spin $s$) and the gravitational field can be formulated as a minimal gauge coupling. 

          \subsection{The coupled system as a minimal gauge coupling}
The formalism we would like to develop is based on the fact that we can identify the degrees of freedom of a massive and spinning relativistic particle evolving in the three dimensional euclidean space ${\mathbb E}^3$ with an element of $G=ISU(2)$. Any element $X \in G$ can be decomposed, by definition, as the semi-direct product $X=(\Lambda,q)$ of a rotation $\Lambda \in SU(2)$ and a translation $q$: we can clearly identify $q$ with the position of a particle in $\mathbb E^3$ and we will show that $\Lambda$ is related to the momentum of the particle\footnote{The momentum $p$ and spin-vector $\sigma$ of a three-dimensional relativistic particle are proportional and related by the expression $\sigma=\frac{s}{m}p$ where $m$ and $s$ are respectively the mass and the spin.}). In that framework, the dynamics between times $t_1$ and $t_2$ of a relativistic particle is defined by the following first order action:
\begin{eqnarray}\label{freeparticleaction}
S_p[X] = \int_{t_1}^{t_2} \!dt  <\chi(m,s)\; , \; X^{-1} \frac{dX}{dt}> \; \text{where} \;  \chi(m,s)=mJ_0 + sP_0 \in \mathfrak g \;.
\end{eqnarray}
The application $<,>:{\mathfrak g} \times {\mathfrak g} \rightarrow \mathbb C$ is a non-degenerate invariant bilinear form on $\mathfrak g$ such that the only non-vanishing product is $<P_a,J_b> = \delta_{ab}$. One can show that the action $S_p$ reproduces the motion of a free massive and spinning relativistic particle: its trajectory is a geodesic of $\mathbb E^3$ and its momentum is conserved \cite{Sousa}.

Before going to the coupled system, it is interesting to note that the free particle action (\ref{freeparticleaction}) is invariant under the global transformation $X \mapsto gX$ for any $g \in G$. The coupling to the gravitational field consists precisely of gauging this global symmetry using a $G$-connection $\omega$. Naturally, the dynamics of the gauge field $\omega$ is given by a Chern-Simons action $S_{CS}[{\omega}]$ of gauge group $G$ defined on a three dimensional manifold ${M}=\Sigma \times [t_1,t_2]$:
\begin{eqnarray}\label{ChernSimonsaction}
S_{CS}[{\omega}] \; = \; \int_{M} d^3x \; \epsilon^{\mu \nu \rho} \left( <{\omega}_\mu,\partial_\nu {\omega}_\rho> + \frac{1}{3} <{\omega}_\mu,[{\omega}_\nu,{\omega}_\rho]> \right) \;.
\end{eqnarray}
A naive minimal gauge coupling introduces ambiguities and therefore it is necessary to consider a regularization. The basic idea\cite{Buffenoir} consists of replacing the point particle by a small loop $\ell$ whose time evolution draws a cylinder ${B}=\ell \times [t_1,t_2]$ in the space-time $M$.

The particle lives on the boundary $B$ and the coupling between the particle and the gravitational field is defined as an integral over $B$:
\begin{eqnarray}\label{actionfortheboundary}
S_c[{\omega},X] \; = \; \frac{1}{2\pi} \int_{B}dt \; d\varphi <\chi(m,s),X^{-1}\frac{dX}{dt} + X^{-1}{\omega}_t X> \;.
\end{eqnarray}
Note that $\varphi \in [0,2\pi]$ is the angular variable parametrizing the loop $\ell$ and the origin $\varphi=0$ is arbitrary. In this integral, $X$ depends only on $t$ and ${\omega}_t$ is the component of the connection in the time direction. 

The dynamics of the gauge field ${\omega}$ is, \`a priori, described by a Chern-Simons action (\ref{ChernSimonsaction}). However, because of the regularization, the space-time admits a boundary $B$ which breaks the gauge invariance. To restore the invariance, first we have to ask the question of what type of symmetry we would like at the boundary. As the parameter $\varphi$ does not have any physical meanning, we impose that gauge transformations are constant along $\ell$. To have a theory invariant under these transformations, the Chern-Simons action has to be supplemented with the boundary term $S_B[{\omega}]  =  \int_{B}dt  d\varphi <{\omega}_t,{\omega}_\varphi>$. It is trivial to verify that the resulting action $S_{CS}[{\omega}]+S_{B}[{\omega}]$ is invariant under the required gauge transformations. Then, the action for the coupled system is given by:
\begin{eqnarray}\label{regularizedcouplingsystem}
S[{\omega},X] \; = \; S_{CS}[{\omega}] \; + \; S_{B}[{\omega}] \; + \; S_c[{\omega},X]\;.
\end{eqnarray}
The generalization to an arbitrary number $N$ of particles is straightforward: we call $\wp_i$ the $i^{th}$ particle of mass $m_i$ and spin $s_i$ whose dynamical variable is denoted $X_i=(\Lambda_i,q_i) \in G$.

          \subsection{Symmetries and Observables}
This section aims to present the main results of the canonical analysis of the action (\ref{regularizedcouplingsystem}). In particular, we obtain the following constraint defined for any $v \in C^{\infty}(\Sigma,\mathfrak g)$:
\begin{eqnarray}\label{regularizedconstraint}
\Phi(v) \; \equiv \; \int_\Sigma d^2x \; \epsilon^{ij}<v,{F}[\omega]_{ij}> + 2 \int_\ell d\varphi <v,{\omega}_\varphi + \frac{1}{2\pi}X\chi(m,s)X^{-1}> \;.
\end{eqnarray}
As a consequence, the connection is flat on the surface $\Sigma$ but its holonomy $H_\ell$ around the loop $\ell$ is given by $H_\ell=X \exp(-\chi(m,s))X^{-1}$. Note that the connection ${\omega}$ is defined everywhere on $\Sigma$ and therefore there is no singularity.

When $v$ is constant along $\ell$, then the constraint (\ref{regularizedconstraint}) is first class and therefore generates gauge transformations. To this constraint, we have to add the constraints that generate internal symmetries associated to the particle: they are parametrized by an element $\varepsilon$ of the Cartan subalgebra of $\mathfrak g$. As a result, infinitesimal gauge transformations on the variables $X$ and on the spatial components of the connection $({\omega}_i)_{i=1,2}$ read:
\begin{eqnarray}\label{gaugetransformationsofcouplingsystem}
\delta X \; = \; -X\varepsilon - v(\ell)X \;\;\;\;\; \text{and} \;\;\;\;\; \delta {\omega}_i \; = \; {D}_iv \;
\end{eqnarray}
where $v(\ell)$ denotes the evaluation of the function $v$ on the loop $\ell$. These results are trivially extended to a system ${P}=\{\wp_1,\cdots,\wp_N\}$ of $N$ particles.

\medskip

In the absence of particles, the theory reduces to a Chern-Simons theory associated to the group $G$ and the physical phase space is known as the moduli space of flat $G$-connections. Classical observables are defined in terms of spin-networks. In the presence of particles, we can define new observables which capture the dynamics of the particle and contain, in particular, the relative positions between particles. To understand this point, let us define the following function:
\begin{eqnarray}\label{configurationoperator}
{O}_\gamma[{\omega},X_i,X_j] \; \equiv \; X_i(t_1)^{-1} U_\gamma[{\omega}] X_j(t_1) \;.
\end{eqnarray}
In this definition, $U_\gamma[{\omega}]$ denotes the holonomy of the connection along the curve $\gamma$ whose starting point and end point are respectively the points of coordinates $\varphi_i \in \ell_i$ and $\varphi_j \in \ell_j$. As the connection $\omega$ is flat, ${O}_\gamma$ depends only on the homotopy class of the path $\gamma$. Moreover, it is clearly invariant under the gauge transformations (\ref{gaugetransformationsofcouplingsystem}) and therefore is an observable. We deal with internal symmetries of the particle by considering initial configuration variables in (\ref{configurationoperator}).

To extract physical informations from ${O}_\gamma[{\omega},X_i,X_j]$, we shall write it in the vectorial representation and we obtain:
\begin{eqnarray}\label{Configurationinthefundamentalrepresentation}
\left(
\begin{array}{cc}
\Lambda_i(t_1)^{-1} h_\gamma[A] \Lambda_j(t_1) & \hspace{0.1cm}\Lambda_i(t_1)^{-1}(h_\gamma[A]q_j(t_1) + q_{\gamma}[e,A] - q_i(t_1)) \\
0 & \hspace{0.1cm}1 \\
\end{array}
\right).
\end{eqnarray}
$h_\gamma[A] \equiv P\exp \int_{\gamma} A$ denotes the holonomy of the spin-connection $A$ along the path $\gamma$ and we have introduced the notation $q_\gamma[e,A] \equiv \int_{\gamma} h_{\gamma<x}[A] e_\mu(x) dx^\mu$ where the path $\gamma<x$ (resp. $\gamma>x$) is the part of $\gamma$ which ends to (resp. starts from) the point $x \in \gamma$. One can see that the translational part ${q}_\gamma(i,j)$ of ${O}_\gamma$ (\ref{Configurationinthefundamentalrepresentation}) is the analogous of the position  of the particle $\wp_j$ in the rest frame of $\wp_i$. Then, the relative distance ${D}_\gamma(i,j)$ is given by the formula:
\begin{eqnarray}\label{distanceoperator}
 {D}_\gamma^2(i,j) \; = \; {q}_\gamma(i,j)^{\dagger} \; {q}_\gamma(i,j) \;.
\end{eqnarray}
This distance operator is a gauge invariant function which depends only on the homotopy class of $\gamma$. As a result, we have constructed a distance operator which is a Dirac observable and therefore should be related to a physical process.

\section{Loop Quantization}
 The following aims at exploring loop quantum gravity techniques as an interesting quantization scheme of the coupled system. In particular, we define the notion of kinematical Hilbert space of the coupled system, we quantize the distance observables and compute their action on particular states.
 
For that purpose, we decompose the first class constraints $\Phi$ (\ref{regularizedconstraint}) into its torsion part $\Phi_T$ and its curvature part $\Phi_C$ defined by:
\begin{eqnarray}\label{separationofconstraints}
\Phi_T(v) \; \equiv \; \Phi(v^a J_a) \;\;\;\; \text{and} \;\;\;\; \Phi_C(v) \; \equiv \; \Phi(v^aP_a)
\end{eqnarray}
for any element $v \in C^{\infty}(\Sigma,\mathbb R^3)$ which is constant on the boundary $\ell$. These constraints respectively generate the rotational and the translational gauge transformations (\ref{gaugetransformationsofcouplingsystem}) on the dynamical variables.

          \subsection{Kinematical Hilbert space}
The strategy will consist of quantizing the theory before implementing the first class constraints. The kinematical Hilbert space ${H}_{kin}(\Sigma,P)$ will be defined as the set of states satisfying all first class constraints but not the curvature constraint $\Phi_C$ endowed with a suitable scalar product. Then, we define the physical Hilbert space $H_{phys}(\Sigma,P)$ from the kinematical one by introducing a projector which implements the constraint $\Phi_C$. Finally, we make use of this projector to define a spin foam model describing dynamics of massive and spinning self-gravitating particles \cite{Noui}.

In the pure gravitational case, kinematical states are defined as gauge invariant cylindrical functions on $\Sigma$. The set of invariant cylindrical functions $\text{Cyl}(\Sigma)$ in naturally endowed with an invariant measure and spin-network states provide it with an orthonormal basis. To generalize this notion to the coupled system, let us note that the action (\ref{freeparticleaction}) of the free relativistic particle $\wp$ admits a ``loop quantization'' scheme which leads to the following formulation for its Hilbert space:
\begin{eqnarray}
{H}(\wp) \; \equiv \; \bigoplus_{j-s \in \mathbb N} \{\stackrel{j}{\pi} \!\!(\Lambda)^l_s  \vert l \in [-j,+j]\} \;.
\end{eqnarray}
In that definition, $\stackrel{j}{\pi}$ denotes the spin $j$-representation of $SU(2)$ and $s$ is the spin of the particle. Therefore, ${H}(\wp)$ is a sub-space of polynomial functions of $SU(2)$ whose Hilbert structure is naturally given by the $SU(2)$ Haar measure. We make use of this formulation to generalize the space of cylindrical functions to the coupled system as follows:
\begin{eqnarray}
\text{Cyl}(\Sigma,{P}) \; \equiv \; \text{Cyl}(\Sigma) \otimes H(\wp_1) \otimes \cdots \otimes H(\wp_N) \;.  
\end{eqnarray} 
The inner products on $\text{Cyl}(\Sigma)$ and on $H(\wp_i)$ naturally endow $\text{Cyl}(\Sigma,{P})$ with a Hilbert structure and we call auxiliary Hilbert space $H_{aux}(\Sigma,P)$ its completion.

The generalized Gauss constraint $\Phi_T$ (\ref{separationofconstraints}) generates $SU(2)$ gauge transformations on the variables of the coupled system. As a result, the kinematical Hilbert space $H_{kin}(\Sigma,P)$ is defined by the Hilbert subspace of gauge invariant functions of $H_{aux}(\Sigma,P)$. At this stage, we can generalize the notion of spin-network states to the coupled system and provide $H_{kin}(\Sigma,P)$ with an orthonormal basis \cite{Noui}. The construction of the physical Hilbert space is presented in \cite{Noui} and is a generalization of the techniques developed in the pure gravitational case \cite{Noui2}.

	  \subsection{Distance operators}
Let us consider a classical (square) distance operator (\ref{distanceoperator}) between the particles $\wp_i$ and $\wp_j$ and associated to an oriented path $\gamma$ linking them. One can see that there is no ambiguity to quantize it and one obtains a quantum operator $\hat{{D}}{}^2_\gamma(i,j)$ acting on ${H}_{kin}(\Sigma,P)$. Let us consider a gauge invariant generalized spin-network state $\Psi_S \in {H}_{kin}(\Sigma,P)$ associated to a graph $\Gamma$ such that there is one and only one edge $\zeta \in {E}_\Gamma$ (colored by $i_\zeta$) which intersects the curve $\gamma$ on only one point $x \not \in \ell_1,\ell_2$. The action of $\hat{{D}}{}^2_\gamma(i,j)$ on such a state is immediate and the result reads\cite{Noui}: 
\begin{eqnarray}
\hat{{D}}{}^2_\gamma(i,j) \; \rhd \; \Psi_S \; =  \; i_\zeta(i_\zeta + 1) \; \Psi \;.
\end{eqnarray}
This result is identical to the usual one obtained in pure loop quantum gravity even if our expression of the distance operator is a priori completely different to the usual one. Nevertheless, the result of the action of the distance operator on a state which intersects several times the path $\gamma$ is not given by a contribution for each edges $\zeta$ intersected. This result is subtil, presented in \cite{Noui} and depends only on the homotopy class of $\gamma$ (manifestation of the diffeomorphism invariance of the distance operator).

\section*{Aknowledgments}
The authors would like to thank Lee Smolin for the kind invitation to the QTS3. This work has been supported in part by NSF grants PHY-0090091 and INT-0307569 and the Eberly Research Funds of Penn State. 

%
%
%
%

\end{document}